\documentclass[prl,preprint,showpacs,groupedaddress]{revtex4}

\usepackage{dcolumn}
\usepackage{graphicx}
\usepackage{amsmath}
\usepackage{bm}

\usepackage[usenames]{color}

\begin{document}

\title{All Optical Switch of Vacuum Rabi Oscillations: \\ The Ultrafast Quantum Eraser}

\author{A. Ridolfo$^1$, R.\ Vilardi$^1$, O.\ Di Stefano$^{1,3}$, S. Portolan$^2$, and S. Savasta$^1$}
\affiliation{$^1$Dipartimento di Fisica della Materia e Ingegneria Elettronica, Universit\`{a} di Messina Salita Sperone 31, I-98166 Messina, Italy} 
\affiliation{$^2$CEA/CNRS/UJF ``Joint team Nanophysics and semiconductors", Institut N\'{e}el-CNRS,
BP 166, 25, rue des Martyrs, 38042 Grenoble Cedex 9, France} 
\affiliation{$^3$Dipartimento di Matematica, Universit\`{a} di Messina Salita Sperone 31, I-98166 Messina, Italy} 
\date{\today}
\begin{abstract}
{We study the all-optical time-control of the strong coupling between a single cascade three-level quantum emitter and a microcavity. We find that only specific arrival-times of the control pulses succeed in switching-off the Rabi oscillations. Depending on the arrival times of control pulses, a variety of exotic non-adiabatic cavity  quantum electrodynamics effects can be observed.
We show that only control pulses with specific arrival times are able to suddenly switch-off and -on  first-order coherence of cavity photons, without affecting their strong coupling population dynamics. Such behavior may be  understood as a manifestation of quantum complementarity. }
\end{abstract}
\maketitle

Control over the interaction between single photons
and individual optical emitters is of great importance
in quantum science and quantum engineering \cite{Monroe}.
In the last years, substantial advances have been made
modifying photon fields around an emitter using high-finesse
optical cavities.
Quantum emitters in a microcavity can absorb
and spontaneously re-emit a photon many times before dissipation
becomes effective and mixed light-matter
eigenmodes arise. 
This strong-coupling regime has been investigated in many systems, ranging from atoms \cite{Haroche, Weis, Reith, Yoshie, Imamoglu,Chiorescu} through excitons in  semiconductor nanostructures, and superconducting qubits.

Recently ultrafast time control of the strong coupling regime has been achieved in a quantum well (QW) waveguide structure \cite{Tredicucci}.
By exploiting an all-optical scheme, the dynamics of a coherent photon population, during sub-cycle switching-on of the coupling, has been observed.
This experiment opens a new domain of non-adiabatic light-matter interaction leading to unconventional QED phenomena \cite{Peropadre}.
In addition, the time-control of vacuum Rabi oscillations can be exploited for the realization of efficient ultrafast all optical switching devices.
However for practical device applications, it should also be possible to switch-off the strong coupling at the same speed.
Gunter et al. \cite{Tredicucci} suggest that the switch-off may be achieved by a pump-dump scheme, where after a strong population inversion in level $| 1 \rangle$  (see Fig.\ 1) is produced by a control pulse (switch-on of the coupling), subsequent ultrafast depopulation is determined by  a second identical control pulse through stimulated emission (switch-off).

In this letter we study the time-control of the strong coupling between a single cascade three-level quantum emitter and a microcavity. The proposed scheme can be applied to a variety of  coupled quantum systems and frequency-ranges of the interacting optical fields.
We find that, while the switch-on  closely resembles the  findings of Ref. \cite{Tredicucci}, a more complex and intriguing behavior is observed when trying to switch-off the strong coupling interaction. Only specific arrival-times of the control pulses succeed in switching-off, hence  posing  severe limitations on the strong coupling full time-control. In particular, strong-coupling is switched-off only when control pulses arrive at  photon population maxima. This implies that the time-width of the switching-off control pulse needs to be short with respect to the frequency of Rabi oscillations.  We also show that, when  switch-off fails, the control pulse may be exploited to suddenly destroy first-order coherence of cavity photons, without affecting their strong coupling population dynamics. Such behavior may be fully understood as a manifestation of quantum complementarity. 
In this case the induced entanglement between the cavity and the quantum emitter enables the latter to store the {\em which way} information on photon paths, hence destroying  coherence according to the quantum complementarity principle \cite{Rempe}. However the loss of coherence is not irreversible, being the which-path detector itself a quantum system (quantum marker) \cite{Herzog}. A further suitable control pulse is able to suddenly erase the {\em which-way} information thus inducing the sudden rebirth of coherence. This scheme thus enables the all optical ultrafast time-control of which-way information in a coupled quantum system.
We also find that control pulses can induce other exotic behaviors, for example  weak and strong coupling dynamics can coexist. Specifically the coherent part of photon population is in the weak coupling regime and at the same time, the incoherent part displays vacuum Rabi oscillations. 

We address an all-optical scheme for time-control of the coupling between the emitter and the cavity analogous to that introduced in Ref.\ \cite{Tredicucci} sketched in Fig.\ 1a. We consider a cascade 3-level quantum emitter as shown in Fig.\ 1b.
A coherent control pulses resonantly and directly drive the
transition $| g \rangle \to | 1 \rangle$, while coherent probe pulses feed the microcavity and their central frequency is chosen to coincide with the resonance frequency of the cavity which in turn is resonantly coupled to the transition $| 1 \rangle \to | 2 \rangle$. As a consequence of polarization selection rules or frequency detuning, we assume that the control and the probe beam do not couple to the transitions $| 1 \rangle \to | 2 \rangle$ and $| g \rangle \to | 1 \rangle$ respectively. The cascade  biexciton $\to$ exciton $\to$ ground-state transitions of a semiconductor QD could be exploited for a solid-state implementation at optical wavelengths of this scheme. 

The master equation for the density operator $\rho$ of the cavity-emitter system can be
written as
\begin{equation}\label{1}
\dot{\varrho}= i \left[\rho,H\right]+\mathcal{L}\rho\, ,
\end{equation}
where the system Hamiltonian reads
\begin{equation}
   H= H_0+ H_I+H_{\rm in},
\end{equation}
with 
\begin{equation}
H_0=						 \sum_{j=1,2}
	  											\omega _j\sigma _{j,j}+
	  											\omega_a {a}^{\dagger} a\, ,
	  											\label{kin}\\	  						
\end{equation}
\begin{equation}
H_I=						g \sigma _{12}  a^\dag + \text{H.c.}\, ,
	  										\label{int}\\
\end{equation}	  											
\begin{equation}
H_{\rm in} = \epsilon_{\rm p}(t)^*a +  \epsilon _{\rm c}(t)^* \sigma _{g,1} + \text{H.c.}
\, .
\label{input}
\end{equation}
Here $a$ denotes the destruction operator for the cavity mode, $\sigma_{\alpha, \beta} \equiv \left| \alpha \rangle \langle \beta \right|$
describes the transition or projection operators involving the levels of the quantum emitter (see also Fig.\ 1).
The Hamiltonian term $H_{\rm in}$ describes both the influence of the driving control pulse $\epsilon _{\rm c}(t)$ directly coupled to the quantum emitter and of the coherent probe pulse $\epsilon _{\rm p}(t)$ feeding the microcavity. We will consider Gaussian ultrafast pulses at frequencies centered at  $\omega_{1}$ and $\omega_{a}= \omega_2 - \omega_1$ respectively.
The Markovian interaction with reservoirs determining the decay rates
for the
QD exciton and the cavity mode respectively, is described by the following Liouvillian terms
\begin{equation}
 \mathcal{L}^{\text{ex}}\varrho = - \frac{1}{2}\sum_{\mu} \left( L^\dag_\mu  L_\mu \rho + \rho  L^\dag_\mu  L_\mu 
 -2  L_\mu \rho  L^\dag_\mu \right)\, ,
\end{equation}
 where the Lindblad operators $\hat L_\mu$ describe the various scattering channels. The decay terms  $2 \to 1$ are described by
 $L_{2 \to 1} = \sqrt{\gamma_2} \left| 1 \rangle \langle 2 \right|$, $ L_{1 \to g} = \sqrt{\gamma_1} \left| g \rangle \langle 1 \right|$. The Limblad operators describing pure dephasing of the levels $\alpha = 1, 2$ are $\sqrt{\gamma^{\rm d}_{\alpha}} \sigma_{\alpha,\alpha}$.

\begin{figure}[t!]  
\includegraphics[height=80 mm]{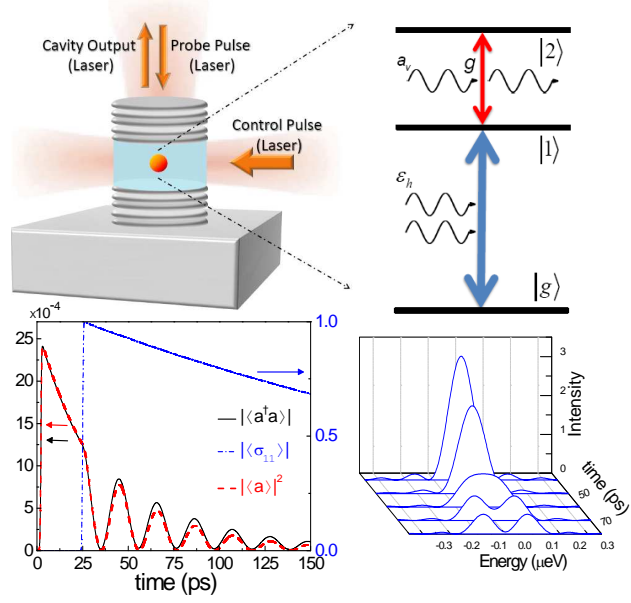}  
\caption{(Color online) All optical control of strong light–matter coupling. (a) Schematic setup to implement the all optical control of the strong coupling regime. (b)  The relevant level structure of the quantum emitter coupled to the resonator.(c)  populations of level $|1 \rangle$ induced by the control-pulse (dash-dotted line), cavity-photon population (continuous line) and its coherent part (dashed). After the arrival of the control-pulse, 
cavity photon populations switch their dynamics from exponential decay to vacuum Rabi oscillations.} 
\end{figure}

Starting from Eq.\ (\ref{1}), we may derive the coupled equations of motion for the cavity-photon and exciton populations, coherences  and higher order correlation functions.
It is known in cavity QED that master equations like (\ref{1}) induce an
open hierarchy of dynamical equations which needs some approximation. A widely adopted
truncation scheme is the one based on the smallness of the excitation density which allows to
truncate with respect to the number of photon number-states to be included. The inclusion
of one-photon states only gives linear optical Bose-like dynamics. Nonlinear optical effects
need the inclusion of multi-photon states. This truncation scheme is particularly suited for the present case where 
weak input pulses feeding the cavity are considered.

We start studying the switch-on of vacuum Rabi oscillations. Fig.\ 1c displays the intracavity photon population (proportional to the output transmitted light intensity). At initial time $t = 0$, the cavity is in the vacuum state and the quantum emitter in its ground state $\left|g \right>$.
At $t = 2$ ps, the cavity is excited by a weak resonant pulse (FWHM = $0.4$ ps). Being levels $| 2 \rangle$ and $| 1 \rangle$ empty, cavity photons are not able to couple to the quantum emitter transitions. As a result cavity photons decay exponentially according to the cavity decay time $1/\gamma_a$.  
At $t = 25$ ps the control pulse $\epsilon_{\rm c}(t)$ resonant with the $| g \rangle \to | 1 \rangle$ transition, with pulse area $\pi$, populates level $| 1 \rangle$.
Almost instantaneously vacuum Rabi oscillations appear. Calculations have been carried out by using $g = 100 \mu$eV, $\gamma_{a} = 20 \mu$eV, $\gamma^{\rm d}= 0$, $\gamma_{xx} = 5 \mu$eV, $\omega_h = 1.3066$eV,  $\omega_3 - 2 \omega_h = -2.27$ meV. 
Comparing the dynamics of the field intensity $\langle a^\dag a \rangle$ and of its coherent part $\left| \langle a \rangle \right|^2$,
despite the absence of pure dephasing, a non-negligible loss of coherence of the photon field after the switch-on is evident. It origins from the 
decay of the effective coupling $\propto \langle \sigma_{11} \rangle$ during the oscillations. We find that decreasing $\gamma_1$ implies a decreasing of pure decoherence.  
The switch on of the coupling  can also be inferred looking at the power spectrum of the coherent part of the light field: $S(T, \omega) = | \int^{T+ \Delta}_{T}  \langle a_v(t) \rangle  \exp{(-i \omega t)} dt |^2$. The ultrafast transition from a single peak centered at the energy of the cavity to th Rabi doublet is evident in Fig.\ 1d.
%
%
\begin{figure}[t]  
\includegraphics[height=65 mm]{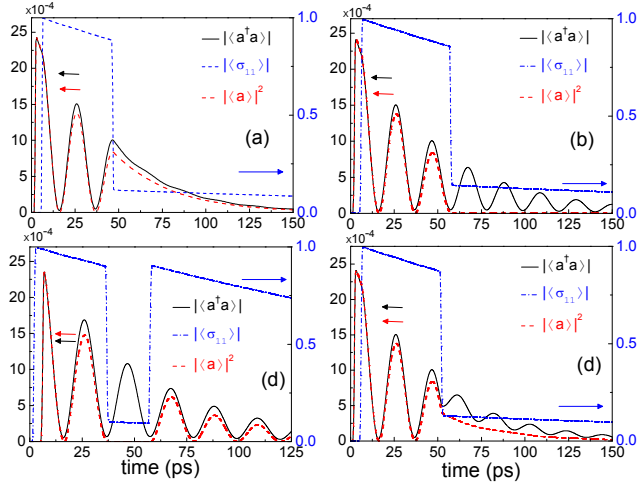}  
\caption{(Color online) Dynamics of cavity-photon populations, its coherent part and populations $\langle \sigma_{11} \rangle$ in presence of two $\pi$ control-pulses (a,b,d), and three control pulses (c). }\label{fig2}
\end{figure}
%
%

Figure \ref{fig2} describes the effect of the arrival of a second control pulse.
We exploit a pair of identical control pulses (each with pulse area $\pi$) which  induce a
strong  population inversion from level $| g \rangle$ to level $| 1 \rangle$ (switch-on of
coupling) followed by ultrafast depopulation of the level $| 1 \rangle$ through stimulated
emission (switch-off). We find that the subsequent dynamics is strongly dependent on the time of the arrival of the control pulse.
If the control pulse depopulates level $| 1 \rangle$ when the photon Rabi oscillations reach the maximum, an almost instantaneous transition from Rabi oscillations to the exponential decay regime can be observed (see Fig.\ 2a).  In addition the coherent part of  the signal closely follows the intensity dynamics.
If the second control pulse excites the system at a cavity-field minimum, switch off fails as shown in Fig.\ 2b. The intensity Rabi oscillations carry on almost undisturbed,
while its previously dominating coherent part suddenly disappears. Such striking dependence of the dynamics on the time of arrival of the control pulse can be explained taking into account the approximate quantum state describing the system when the cavity is feeded by a low intensity pulse (probe).
The general quantum state after the arrival of the probe and of the first control pulse can be developed as

\begin{equation}|
	\psi(t) \rangle = \alpha(t) | 1 \rangle | 1 \rangle + \beta(t)| 0 \rangle | 1 \rangle + \gamma(t)| 0 \rangle | 2 \rangle\, ,
\label{psi1}
\end{equation}
 where the first ket describes the photon state and the second the quantum emitter state. At a cavity maximum, $\alpha(t_M)$ is maximum and $\gamma(t_M) \simeq 0$. After the arrival of the second control pulse, the state suddenly becomes
$
| \psi_M(t^+_M) \rangle = [\alpha(t^+_M) | 1 \rangle  + \beta(t^+_M)| 0 \rangle ]| g \rangle
$.
Now, being level $| 1 \rangle$ unoccupied, during the free evolution ($t > t_M$) the absorption of a light quantum by the QE associated to the transition $| 1 \rangle_{\rm QE} \to | 2 \rangle_{\rm QE}$
is not allowed and the state $| \psi_M(t) \rangle$ follows the noninteracting cavity dynamics,
\begin{equation}
| \psi_M(t) \rangle = [\alpha(t) | 1 \rangle  + \beta(t)| 0 \rangle ]| g \rangle\, ,
\label{state_M}
\end{equation}
implying the exponential decay of $\langle a^\dag a \rangle$ shown in Fig.\ 2a. The signal preserves its coherence ($\langle \psi | a | \psi \rangle \neq 0$) in agreement with numerical calculations in Fig.\ 2a.

At a cavity minimum, $\alpha(t_m) \simeq 0$  and $\gamma(t_m)$ is maximum. After the arrival of the second control pulse implying $| 1 \rangle \to | g \rangle$, the state $| \psi(t^-_m) \rangle$ suddenly becomes,  $| \psi_m(t^+_m) \rangle = \beta(t^+_m) | 0 \rangle | g \rangle + \gamma(t^+_m)| 0 \rangle | 2 \rangle$. During the free evolution ($t > t_m$) the emission of a light quantum by the QE associated to the transition $| 2 \rangle_{\rm QE} \to | 1 \rangle_{\rm QE}$
is allowed and the state  can evolve as 
\begin{equation}
| \psi_m(t) \rangle = \alpha(t)| 1 \rangle | 1 \rangle + \beta(t) | 0 \rangle | g \rangle + \gamma(t)| 0 \rangle | 2 \rangle\, .
\label{state_m}
\end{equation}
According to the dynamics of this state, the cavity-photon population preserves its oscillations following the dynamics of spontaneous emission of a two-level QE in the strong coupling regime, with the upper level as initial state: switch-off hence fails. On the contrary, it is straightforward to obtain from Eq.\ (\ref{state_m}) $\langle \psi_m | a | \psi_m \rangle =0$. This is the case depicted in figure fig2b. This result is a direct consequence of the entanglement  created between the quantum emitter and the cavity. Eq.\ (\ref{state_m}) shows that the cavity state is correlated with the quantum emitter state, in contrast to the factorized state
$ |  \psi_M(t) \rangle$. This entanglement is the crucial point for the storage of information which erases coherence according to the complementarity principle.

The coherence $\langle a \rangle$ can be measured by homodyne detection making the signal escaping the cavity interfering with a portion of the input laser beam used to excite the cavity.
This measurement scheme can also be exploited to understand this peculiar behavior in terms of the quantum complementarity principle. When the system state is (\ref{state_M}), the detection of a photon cannot tell us if the photon comes from the cavity or the laser. On the contrary  the correlated state (\ref{state_m})  stores the which-way information in the quantum emitter state: if the detected photon (and only if) comes from the cavity, the quantum emitter will be in state $| 1 \rangle_{\rm QE}$. The which-way information can be read out later by performing a measurement of the QE state. The result of this measurement reveals which way the detected photon took. According to the fundamental quantum complementarity principle coherence is  suddenly destroyed.
%
%
\begin{figure}[t]  
\includegraphics[height=45 mm]{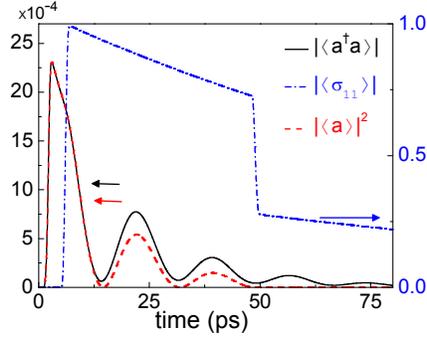}  
\caption{(Color online) Switch-off of photonic coherence in presence of pure dephasing}\label{fig3}
\end{figure}
%
%
The lost (or more appropriately hidden) coherence can be promptly recovered by sending a successive $\pi$ control pulse with arrival time at a photon-population minimum (see Fig.\ \ref{fig2}c).  Results presented in Fig.\ \ref{fig2}c demonstrates the possibility of complete switching control over the photonic coherence.
This behaviour poses severe limitations on the control strategies because only specific arrival times can be employed. Indeed, when the control pulse is sent in the middle between the photon maximum and minimum (Fig.\ 2d), the  photon population  still displays Rabi oscillations (although less pronounced), while the coherent part of the transmitted light displays an exponential decay typical of the weak-coupling regime.

Fig.\ \ref{fig3} describes the switch-off of photon coherence in presence of pure dephasing. Here we consider calculations on a state of the art solid-state microcavity coupled to a single quantum dot.
The results put forward the robustness of the switching against pure dephasing and larger decay rates.


\end{document}